\documentclass[conference]{IEEEtran}
\usepackage{cite}
\ifCLASSINFOpdf
\else
\fi

\usepackage{graphicx}
\usepackage{subfigure}
\usepackage{multicol}
\usepackage{amsmath}

\newcommand{\xhdr}[1]{\vspace{1.7mm}\noindent{{\bf #1.}}}

\begin{document}
\title{\vspace{4mm}Driver Identification Using Automobile Sensor Data from a Single Turn}

\author{
    \IEEEauthorblockN{David Hallac\IEEEauthorrefmark{1}, Abhijit Sharang\IEEEauthorrefmark{1}, Rainer Stahlmann\IEEEauthorrefmark{3}, Andreas Lamprecht\IEEEauthorrefmark{3}, Markus Huber\IEEEauthorrefmark{2}, Martin Roehder\IEEEauthorrefmark{2}, \\
    Rok Sosi\v{c}\IEEEauthorrefmark{1}, Jure Leskovec\IEEEauthorrefmark{1}}
    \vspace{-3mm}
    \and
    \IEEEauthorblockA{\IEEEauthorrefmark{1}Stanford University
    \\ \small\{hallac, abhisg, rok, jure\}@stanford.edu}
    \and
    \IEEEauthorblockA{\IEEEauthorrefmark{2}Volkswagen Electronics Research Laboratory% Group of America, Inc. %- Electronics Research Laboratory
    \\ \small \{markus.huber, martin.roehder\}@vw.com}
    \and
    \IEEEauthorblockA{\IEEEauthorrefmark{3}AUDI AG
    \\ \small\{rainer.stahlmann, andreas.lamprecht\}@audi.de}
}

\maketitle

\begin{abstract}
As automotive electronics continue to advance, cars are becoming more and more reliant on sensors to perform everyday driving operations. These sensors are omnipresent and help the car navigate, reduce accidents, and provide comfortable rides. However, they can also be used to learn about the drivers themselves. 
In this paper, we propose a method to predict, from sensor data collected at a single turn, the identity of a driver out of a given set of individuals. We cast the problem in terms of time series classification, where our dataset contains sensor readings at one turn, repeated several times by multiple drivers. 
We build a classifier to find unique patterns in each individual's driving style, which are visible in the data even on such a short road segment. 
To test our approach, we analyze a new dataset collected by AUDI AG and Audi Electronics Venture, where a fleet of test vehicles was equipped with automotive data loggers storing all sensor readings on real roads. 
We show that turns are particularly well-suited for detecting variations across drivers, especially when compared to straightaways.
We then focus on the 12 most frequently made turns in the dataset, which include rural, urban, highway on-ramps, and more, obtaining accurate identification results and learning useful insights about driver behavior in a variety of settings.

\end{abstract}


% \fi
\IEEEpeerreviewmaketitle

\section{Introduction}

Modern automobiles contain hundreds or even thousands of sensors \cite{SRM:95}, measuring everything from fuel level to the current slope of the road. These sensors are often used to help the car adapt to the current environment (i.e., turning up the heat when it is cold outside, or stopping with the emergency brake when about to hit an object). However, it is also possible to use sensors to learn about the drivers themselves. With these sensors, one can understand different driving styles \cite{VLZ:14}, uncover patterns \cite{SSH:12}, and even detect distracted or impaired drivers \cite{G:98}. To help with these applications and more, it is important to be able to determine who amongst a small set of potential drivers is currently behind the wheel. Everyone drives differently, and the hope is to leverage these differences to find a unique ``signature'' for each driver, which can be anything from how hard the gas pedal gets pressed to small micro-adjustments in the steering wheel angle when turning.

Driver identification is especially useful if user identity can be inferred from just a short snippet of sensor data, such as a single turn.
Analyzing driver behavior on short road segments, rather than over long drives, opens up many new applications.
For example, it would allow the vehicle to recognize, as soon as the car turns out of the driveway, which member of a household is currently driving it. The vehicle could then automatically adjust the settings to fit the driver's preferences (temperature, radio station, mirror placement, etc.). Furthermore, correctly identifying drivers would allow cars to build driver profiles. Vehicles would be able to determine if certain drivers are more aggressive than others, or if some prefer back-road routes to the main roads when navigating. Analyzing behavior at such small granularity would also allow for profiles of each segment of road \cite{I:15}, for example warning the driver to be careful if the car had previously needed to use the emergency brake at an upcoming intersection.
Additionally, this type of analysis could detect changes in driver behavior throughout a drive, such as when an individual uses a handheld cellphone and becomes distracted, since this would manifest itself as a sudden shift in the driving patterns. Note that all of these applications can be implemented locally, without the need for global coordination between different cars. This is imperative because it keeps driver information private, and no data ever has to leave the car. 

For many of these applications, it is preferable or sometimes even necessary to infer the driver's identity from just a small road segment. Any algorithm must therefore find sufficient data in the sensor readings from this very short time interval. This is why turns are particularly well-suited for this type of classification. At a turn, a driver needs to slow down, turn on the blinker, rotate the wheel, determine the car's trajectory and curvature, and then accelerate again, all in a matter of seconds. In other situations, such as straightaways, there are fewer distinct driver actions captured by the sensors. As a result, more complicated prediction models may overfit to the noise, since there may not be enough relevant information to discern between individuals. Of course, for both turns and straightaways, there is a risk that the same driver may behave differently when repeating the segment of road, due to anything from crossing pedestrians to driver mood. However, this is why it is necessary to focus on areas where algorithms yield high prediction accuracy, since these are the regions that allow for the many direct benefits of driver identification.

In this paper, we analyze an anonymized dataset, produced by AUDI AG and Audi Electronics Venture, where drivers were asked to drive modified Audi vehicles equipped with data loggers to store all their sensor readings. The contractors drove the cars on real roads in and around Ingolstadt, Germany. We develop a classification algorithm to predict driver identity from a short segment of road, such as single turn. Our dataset then consists of snippets of sensor readings at this given location, each session approximately 8-10 seconds long and labeled with the driver ID. We build a robust classifier to identify who, out of a set of potential candidates, is operating the vehicle. We test our approach using the Audi data, where we empirically discover that turns are better able to distinguish unique driver patterns than straightaways are. Focusing on turns but ensuring that we are not overfitting our model to one turn type in particular, we independently analyze each of the 12 most frequent turns in the dataset. Our algorithm yields promising results, with average single-turn prediction accuracy of 76.9\% for two-driver and 50.1\% for five-driver classification. Furthermore, we discuss several new insights, such as the fact that drivers appear to be easier to classify in rural areas than urban ones. This is likely because rural locations are more consistent from session to session, since there are limited potential distractions.

The main contributions of this paper are as follows:
\begin{itemize}
\item We collect and analyze a new dataset containing sensor data from Audi cars on real roads.
\item We develop an algorithm to identify a driver, out of a given set of individuals, using only a short segment of vehicular sensor data, such as a single turn.
\item We test our method on the Audi dataset, showing how it is well-suited for single-turn driver identification in various different settings.
\end{itemize}

\xhdr{Related Work}
This work relates to recent research in several different fields. Timestamped sensor data has been used for predicting user identity from walking patterns \cite{MLVMA:05, SYJYX:11}, touch-based biometrics \cite{BZLHW:13, MESR:14}, and head-mounted displays \cite{RWSV:15}. In automobiles, such sensors have been used in understanding driving behavior. It has long been known that individuals have different driving preferences \cite{F:98}. More recently, researchers have built models using sensors to recognize driving styles \cite{DGG:14, VLZ:14}, classify driving events \cite{JT:11, SSH:12}, and detect drowsy drivers \cite{G:98}. Spectral analysis has also been shown to perform well in modeling driver behavior \cite{M:07, OWMIT:05}, which we take advantage of in our classifier.

Van Ly et al. \cite{VMT:13} and Miyajima et al. \cite{M:07} both used inertial sensors to identify drivers from their driving patterns, similar to our approach. Van Ly et al. treat a driving session as a series of events, each of which they classify independently. However, their model does not account for the temporal or spectral aspects of the time series, instead focusing on characteristic statistics in their feature vectors. Our work includes additional sensors, different features, and a more robust method of classification, which greatly improve results. We also experiment on a much larger dataset and test our algorithm on a diverse set of turn types. Miyajima et al. use a Gaussian Mixture Model to solve the driver prediction problem. While they achieve very accurate results, they focus on driver identification over long time series and found that their algorithm's accuracy drops when identifying shorter sessions. Whereas their shortest test was on a 1-minute long dataset, our experiments attempt to identify drivers from a single turn, which only lasts for a few seconds. Both their method and ours incorporate spectral features, though we use slightly different frequency information and apply a different classifier on top of these features. Our work builds off this previous research, developing a method for driver identification that is well-suited for classification from a short snippet of data, allowing for new applications and potential use cases of this work.

\xhdr{Outline}
The rest of this paper is structured as follows. In Section II, we introduce and analyze the novel dataset. In Section III, we explain our classification algorithm. We then define our evaluation criteria in Section IV. In Section V, we present and discuss the results. Finally, we conclude and list potential directions for future work in Section VI.

\section{Dataset Description}
To perform our analysis, we test our results on a new dataset collected by AUDI AG and Audi Electronics Venture. 

\xhdr{Data Acquisition}
To generate this data, contractors were hired to drive retrofitted Audi A3 vehicles, modified to store all sensor readings from the different electronic systems in the car. In order to capture all data, an offline device that writes data to a standard SD card was used. The data rate was approximately 4GB per day per vehicle. A capture device stored a timestamped dump of the CAN-bus communication. At the end of each day, the drivers placed the SD cards in a copy-station to start the parallelized process of moving all data to a central data warehouse. Here, the raw communication data was stored in a PostgreSQL database. 
A drastic reduction of the size of the dataset was achieved by implementing a change detection on each single sensor value, so data was stored whenever it changed, rather than at synchronous intervals.

\begin{figure}[]%h puts in-line
    \centering
    \includegraphics[width=0.99\linewidth]{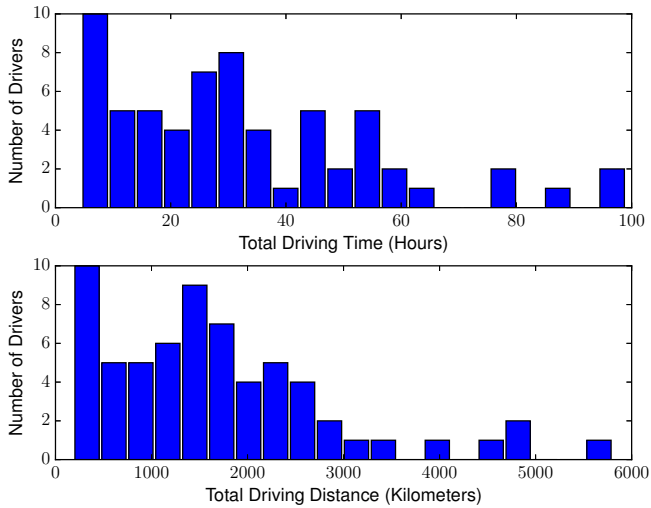}
    \vspace{-5mm}
    \caption{Total driving time and distance for each of the 64 drivers.}
    \vspace{-5mm}
    \label{fig:session}
    \vspace{-1mm}
\end{figure}

\xhdr{Aggregate Statistics}
The dataset consists of 10 cars and 64 total drivers\footnote{No personal information about the drivers was used in this study.}, containing 2,098 total hours of driving and covering 110,023 kilometers. Each morning, drivers would pick up their car from the same Audi parking lot in Ingolstadt. In total, there were 1,889 ``sessions'', continuous time intervals where the driver was actively driving, which were broken up intermittently whenever the drivers would stop for lunch, gas, etc. As such, a typical day would consist of 3-4 ``sessions'', though this number varied widely. Histograms of total driving time and total driving distance for each of the 64 drivers are displayed in Figure \ref{fig:session}. 

\begin{figure}[]%!t?
\centering 
  \subfigure[Initial turn after picking up car.]{\label{fig:turn1}\includegraphics[width=0.493\linewidth]{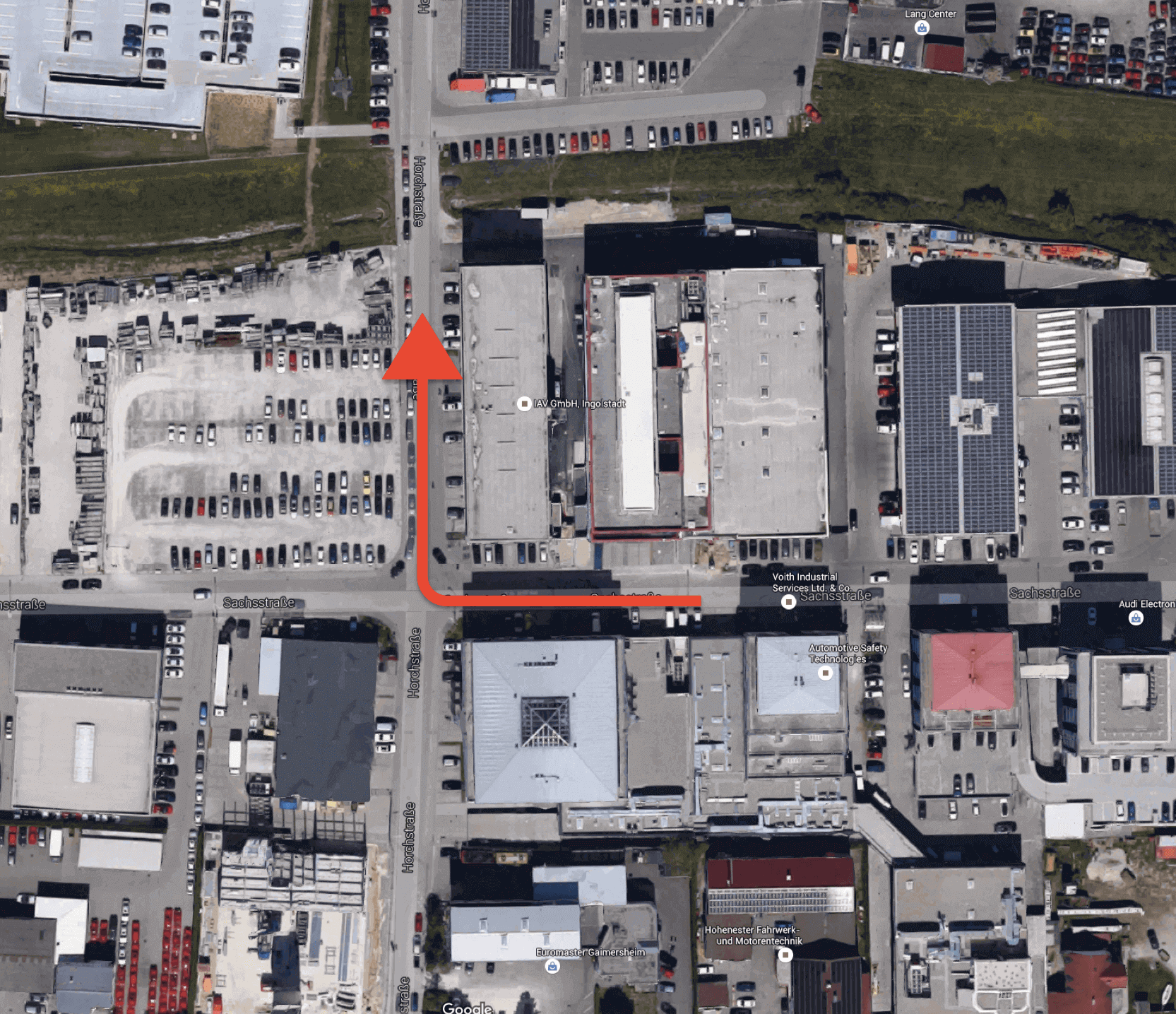}}
  \subfigure[Merge onto busy road.]{\label{fig:turn2}\includegraphics[width=0.493\linewidth]{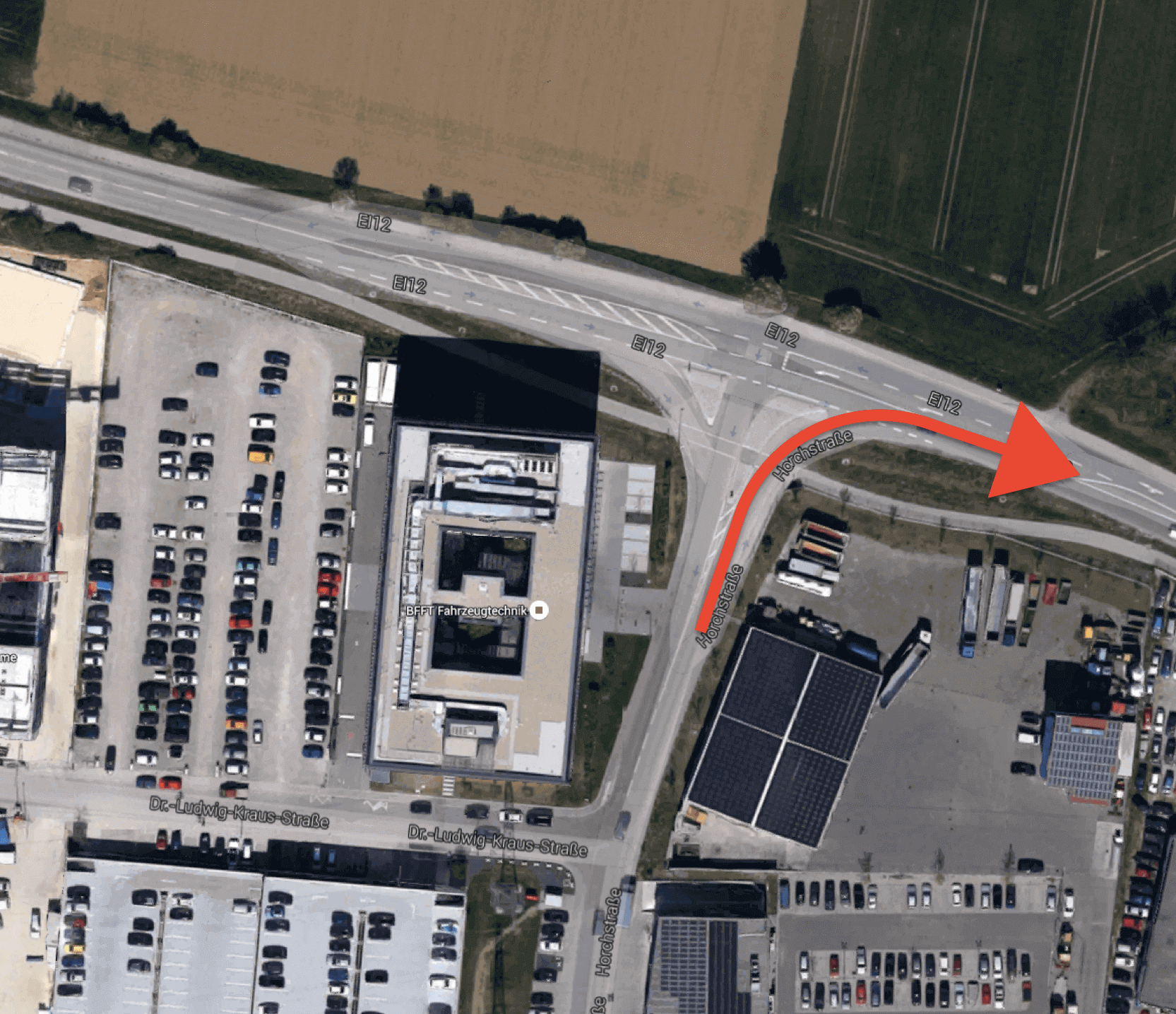}}
  \subfigure[Turn in rural area.] {\label{fig:turn3}\includegraphics[width=0.493\linewidth]{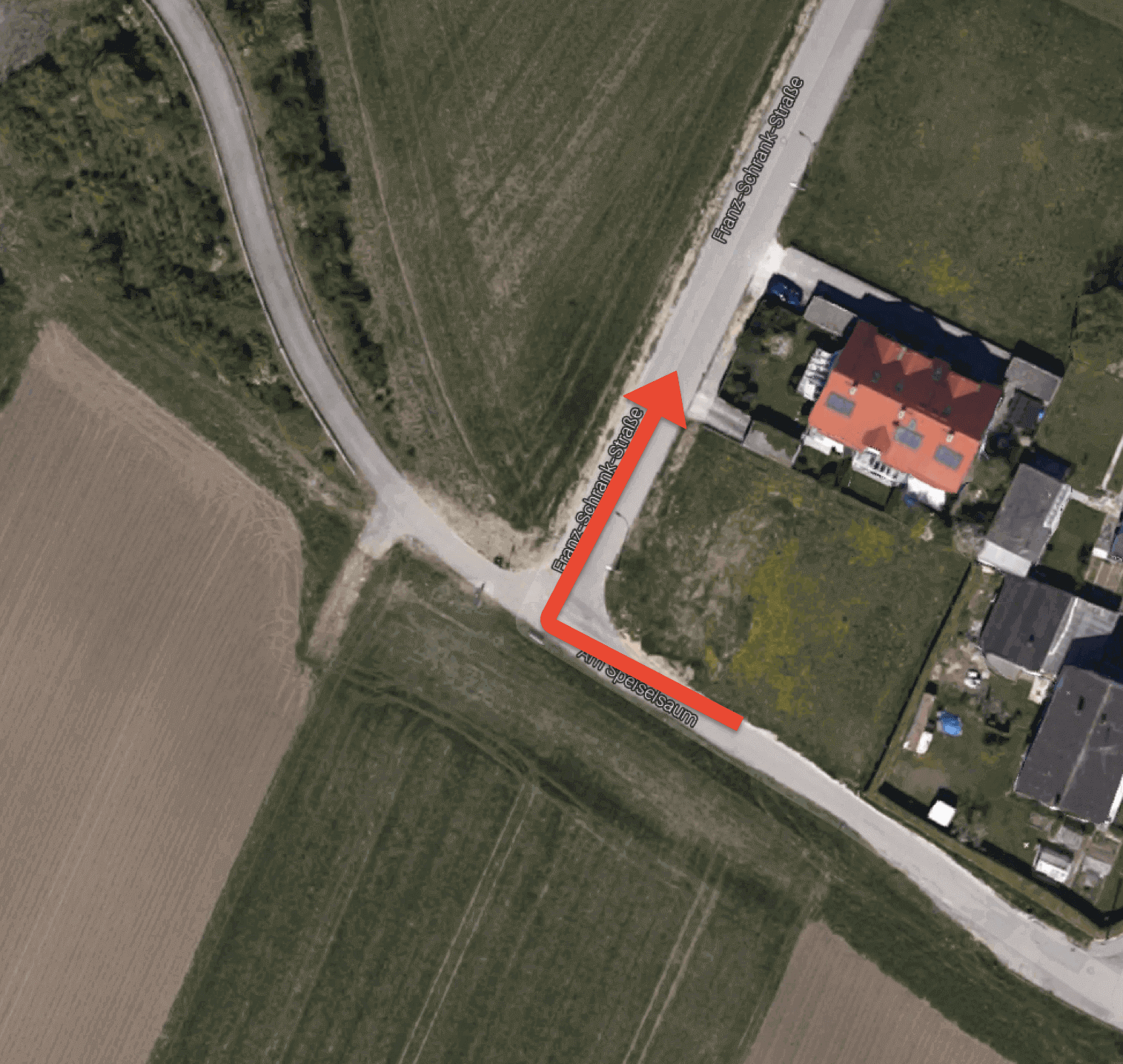}}
  \subfigure[Highway on-ramp.] {\label{fig:turn4}\includegraphics[width=0.493\linewidth]{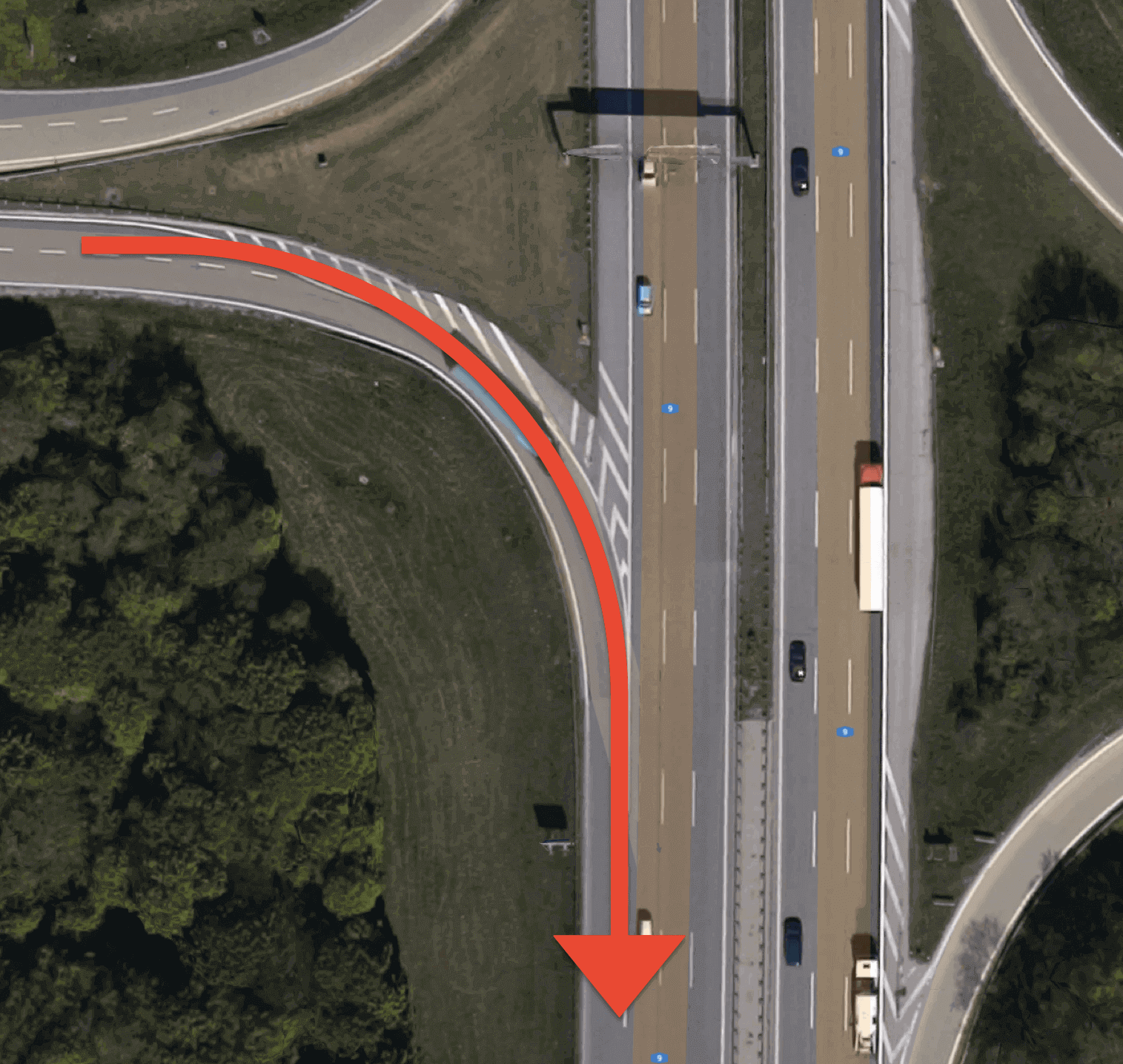}}
  \vspace{-3mm}
   \caption{Four of the twelve most common turns in the dataset.} 
   \vspace{-1mm}
   \label{fig:turns}
\end{figure}

\xhdr{Defining a Turn}
We consider a \emph{turn} as a location where sensors satisfy the following conditions:
\begin{enumerate}
\item Change in heading direction of at least 70 degrees.
\item Total duration lasts less than 10 seconds.
\item ``Stable'' heading for at least 5 seconds before the turn.
\end{enumerate}
These constraints ensure that we limit the number of false positives, for example due to curvy roads or in parking lots. Note that the most frequent turn in the dataset is actually the very first turn outside of the Audi parking lot where all the contractors picked up the cars. This aligns with our goal to identify individuals right at the beginning of their drive. For this paper, we test our algorithm on each of the 12 most common turns in the dataset, so that we avoid overfitting to one turn in particular. We manually examine the selected latitude/longitude data to confirm that they are not false positives, and then label each with a ``turn type''. These 12 turns contain both right and left turns from a diverse set of locations, including busy city roads, empty rural areas, and ramps on/off highways. We include images\footnote{Turn images come from Google Maps, \copyright 2016.} of several of the top turns in Figure \ref{fig:turns}.

\begin{figure}[]
    \centering
    \includegraphics[width=0.99\linewidth]{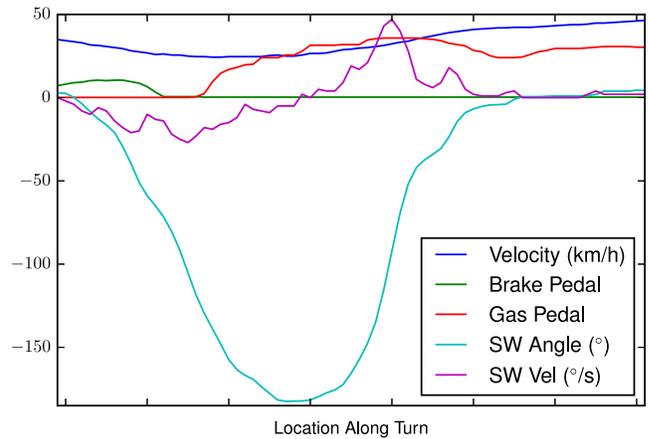}
    \vspace{-5mm}
    \caption{Several key sensors for a single session of a ``typical''  turn.}
    \vspace{-1mm}
    \label{fig:session1}
\end{figure}

\begin{figure}[]
    \centering
    \includegraphics[width=0.99\linewidth]{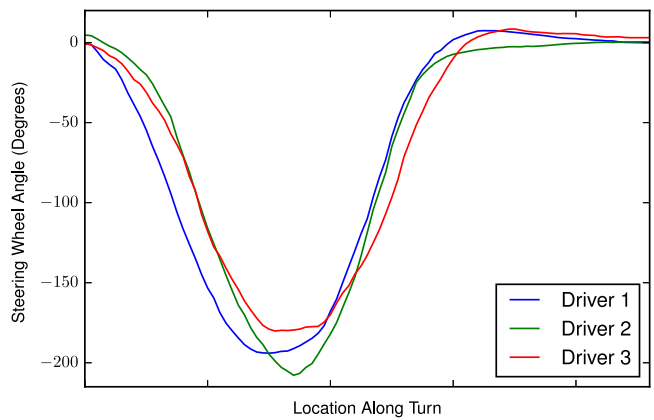} 
    \vspace{-5mm}
    \caption{Average steering wheel angle for three drivers at a single turn.}
    \vspace{-5mm}
    \label{fig:session2}
\end{figure}

It is important to understand how, at a single turn, sensor data can be used to distinguish between drivers. For example, consider the turn displayed in Figure \ref{fig:turn3}. We first plot several sensor readings from a single session of this turn in Figure \ref{fig:session1}. This shows how a driver will slow down, turn the steering wheel (SW), make the turn, then press the gas pedal to speed back up. In Figure \ref{fig:session2}, we compare the average steering wheel angle across all sessions of this same turn for three drivers, each of whom made the turn between 12 and 18 times. Note that driver 1 started turning the wheel earliest, driver 2 turned the wheel farthest at the peak of the turn, and driver 3 kept the wheel turned latest. These are the types of distinguishing characteristics, unique to each person, that we hope to discover with our classification algorithm.

\section{Classification Algorithm}
Consider a single turn, or any short segment of road, driven many times by several drivers. We aim to build a classifier that, given a new session completing this turn, can predict which individual was driving the car. We analyze a 150-foot radius around a given centerpoint, based on GPS readings. In this interval, which typically lasts around 10 seconds, we look at 12 different sensor readings every 0.1 seconds, corresponding to 12 of the signals that are most directly impacted by the driver's actions. Note that the distance between these readings depends on the current speed of the car. At the slow velocities that turns typically occur at, the 0.1s intervals are usually 1-3 feet apart. The 12 signals we use, not counting GPS, are:
    \begin{multicols}{2}
    \begin{itemize}
        \item Steering wheel angle
        \item Steering velocity
        \item Steering acceleration
        \item Vehicle velocity
        \item Vehicle heading
	\item Engine RPM
        \item Gas pedal position
        \item Brake pedal position
        \item Forward acceleration
        \item Lateral acceleration
        \item Torque 
        \item Throttle position
    \end{itemize}
    \end{multicols}

\begin{figure}[]%h puts in-line
    \centering
    \includegraphics[width=0.81\linewidth]{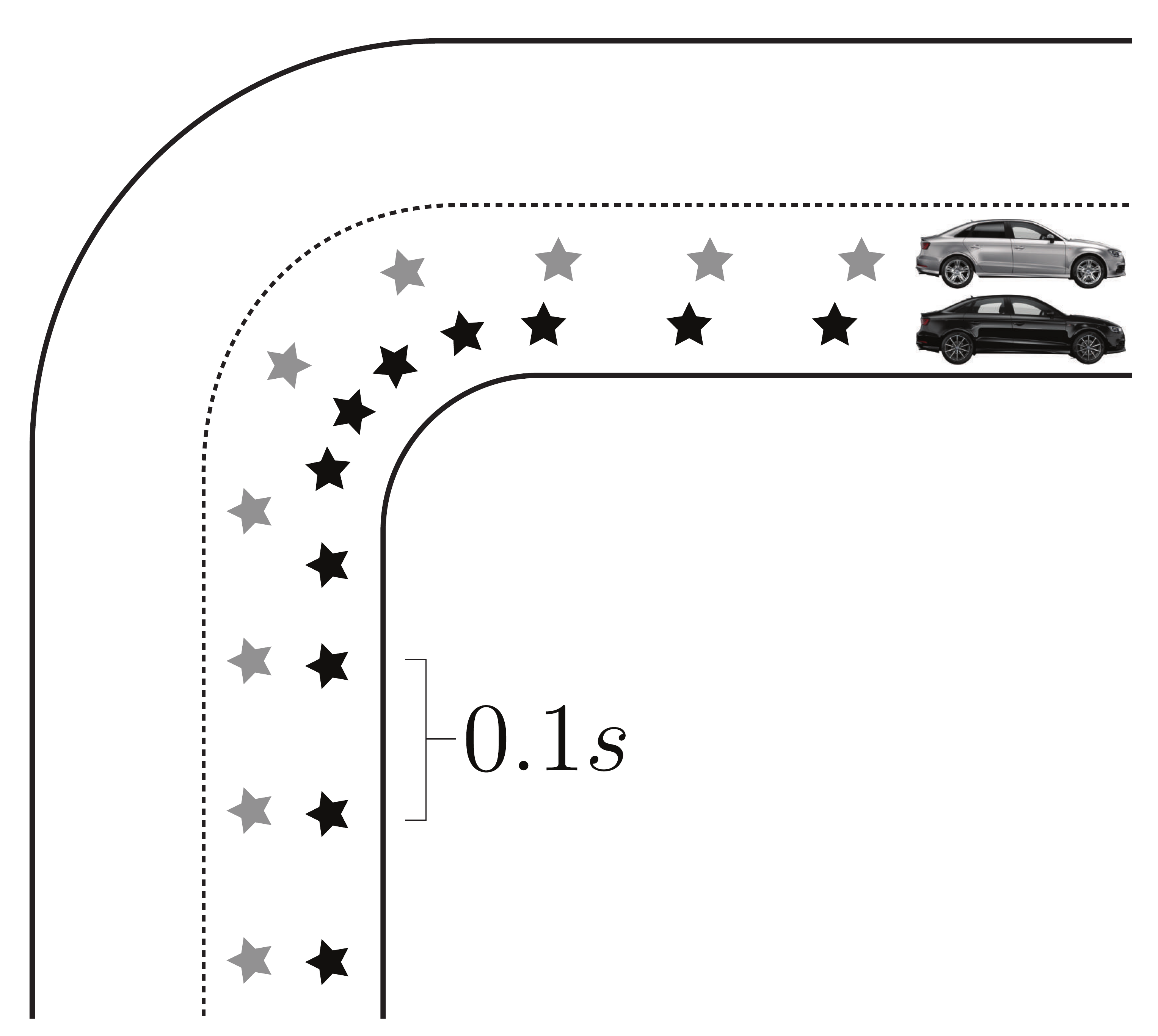}
    \vspace{-1mm}
    \caption{Two cars making the same turn can have their sensor readings become ``misaligned'' if they travel at different velocities.}
    \vspace{-5mm}
    \label{fig:alignment}
\end{figure}

\xhdr{Alignment}
While the sensor values are observed at synchronous intervals, it is necessary to align the readings across different sessions. This is because different sessions of the same turn take different amounts of time, as visualized in Figure \ref{fig:alignment}. We want to compare behavior at a certain location (i.e., 10 feet before the intersection) rather than at a certain time (i.e., 3 seconds after the turn starts), so the raw time-series is not sufficient.

To align by location, we first interpolate the latitude and longitude readings, which arrive at 1Hz, up to the 10Hz sampling frequency of the sensors. Next, we choose one session as our baseline, which we use to align all the other sessions. While our results are relatively robust to the selection of this baseline, we want this session to be as ``smooth'' in velocity as possible. We do so by choosing the session that minimizes 
$$\frac{1}{\sqrt{K}} \sum_{i=1}^{K-1} (V_{i+1} - V_i)^2,$$ 
where $V_i$ is the velocity at state $i$ and $K$ is the number of 0.1-second states in the session. With this baseline selected, we treat its latitude/longitude readings every 0.1s as the ``ground truth'' locations. For every other session, we estimate the sensor readings at these ground truth points. Note that we have not observed the values of the other sessions at the exact ground truth locations, since different sessions can be misaligned, like in Figure \ref{fig:alignment}. Instead, we estimate the value as a weighted average of the two nearest readings (one slightly before, one slightly after). Now, we have an aligned series of $K \approx 100$ lat/long locations, along with the (estimated) sensor readings at these points for \emph{every} session where the turn was made. Therefore, we have a $K$-dimensional vector for each of the 12 sensors. This yields a set of $K \times 12$ data matrices, one per session, each labeled with a driver ID.

\xhdr{Classification Features}
Even with this alignment scheme, a classifier using just the raw data would not perform well, as it does not account for the temporal dependencies at each reading. For example, the velocity at state $i$ is heavily correlated with the velocities at states $i-1$ and $i+1$. Therefore, treating each feature independently can lead to skewed results. Instead, we transform the raw data into two types of features better-suited for time series classification. For every one of the 12 sensors, we generate:

\begin{itemize}
\item \textit{Simple Features} --- Characteristic statistics such as higher order moments provide a low-rank description of the dataset, and have been shown to yield useful insights for temporal analysis with minimal additional computational complexity \cite{WSHA:04}. For each sensor, we record the mean, standard deviation, skew, kurtosis, minimum, maximum, and autocorrelation, representing the $K$-dimensional data as 7 easy-to-calculate points.
\item \textit{Complex Features} --- Additionally, we aim to capture the spectral component of the features. We do so by taking the discrete wavelet transform (DWT) \cite{JC:01} of each sensor's readings, which provides relevant information in both the time and frequency domains. For simplicity, we choose to use the Haar wavelet, one of the most commonly-used wavelets for non-smooth functions \cite{M:08}. This yields a new $K$-dimensional DWT vector for each sensor, on which we then perform Principal Component Analysis (PCA) down to 5 dimensions.
\end{itemize}

With these 12 features per sensor --- 7 simple and 5 complex --- we have transformed the data into a low-dimensional representation, which helps prevent overfitting and allows for computationally efficient algorithms. To solve the driver prediction problem, we build a random forest classifier with this data \cite{B:01}. This method uses training data to build a series of decision trees, which are then used to identify the driver whenever a new session comes in.

\section{Evaluation Criteria}

We evaluate our method on Audi's real-world dataset. We examine the 12 most common turns, as determined by our criteria in Section 2, so that we do not overfit our model to one specific situation. We analyze each turn independently; that is, we retrain our classifier at each turn, without incorporating any information from any of the other turns. For each of the 12 turns, we take the $n$ drivers with the most sessions, varying $n$ from 2 to 5. 
We take a balanced training set, where every driver has the same number of sessions --- if one driver has made the turn more times than another, we drop sessions in chronological order. Finally, we evaluate our prediction accuracy for different turns and different values of $n$. Note that we are not ``throwing away'' any bad data. This includes sessions which may be impeded, for example if a pedestrian crosses in front of the car or if there is a slow driver ahead, since we hope our classifier is robust enough to give accurate predictions in \emph{all} cases, not just ideal turns. We perform stratified $k$-fold cross validation with reshuffling, where $k$ equals the number of sessions per driver for that turn, and validate our method by measuring identification accuracy.

\begin{table}
    \caption{Prediction accuracy for each of the top 12 turns.}
        \vspace{-1mm}
    \scalebox{0.82}{
\begin{tabular}{ c | c | c | c | c | c }
  Turn & Turn & \multicolumn{4}{c}{Prediction Accuracy (\# of sessions per driver)} \\ 
  ID & Type & $n$ = 2 & $n$ = 3 & $n$ = 4 & $n$ = 5  \\
  \hline			
  1 & Urban, non-major road & 72.0\% (34) & 58.7\% (31) & 49.3\% (30) & 41.0\% (29) \\ 
  2 & Urban, non-major road & 75.0\% (24) & 64.0\% (22) & 64.0\% (20) & 54.6\% (19) \\ 
  3 & Merge onto busy road & 69.5\% (24) & 48.3\% (19) & 39.0\% (17) & 33.4\% (12) \\ 
  4 & Merge off busy road & 62.0\% (17) & 56.3\% (17) & 50.3\% (17) & 45.6\% (17) \\ 
  5 & Merge onto highway & 55.0\% (21) & 52.3\% (15) & 57.5\% (12) & 48.2\% (4) \\ 
  6 & Merge onto highway & 79.5\% (17) & 57.0\% (17) & 60.5\% (16) & 49.5\% (16) \\ 
  7 & Urban, major road & 67.5\% (17) & 56.0\% (16) & 60.5\% (13) & 56.8\% (8) \\ 
  8 & Urban, major road & 87.0\% (16) & 51.3\% (12) & 47.0\% (8) & 33.2\% (6) \\ 
  9 & Turn onto bridge & 83.5\% (18) & 60.0\% (15) & 35.0\% (11) & 41.0\% (11) \\ 
  10 & Rural turn & 82.5\% (17) & 74.0\% (15) & 65.8\% (14) & 62.4\% (13) \\
  11 & Rural turn & 93.5\% (16) & 56.7\% (15) & 63.2\% (13) & 70.6\% (12) \\ 
  12 & Rural turn & 96.0\% (13) & 77.7\% (12) & 70.2\% (11) & 64.8\% (11) \\
  \hline
  & Urban/highway average & 72.3\% & 60.0\% & 51.5\% & 44.8\% \\
  & Rural average & 90.7\% & 69.4\% & 66.4\% & 65.9\% \\
  & Average across all turns & \textbf{76.9\%} & \textbf{59.4\%} & \textbf{55.2\%} & \textbf{50.1\%} 
      \vspace{-2mm}
\end{tabular}
}
    \label{results}
\end{table}

\begin{table}
	\centering
    \caption{Top five (out of 12 total) most important sensors for each of the 12 turns in the dataset.}
    \vspace{-2mm}
\begin{tabular}{ c | c | c | c | c | c  }
  Turn & \multicolumn{5}{c}{Most Important Sensors in Random Forest Classifier} \\
  ID & 1st & 2nd & 3rd & 4th & 5th \\
  \hline			
  1 & SW Vel & Speed & Gas & SW Acc & Brake \\ 
  2 & SW Acc & X-Acc & Brake & Throttle & Heading\\ 
  3 & RPM & Throttle & Brake & Speed & SW Acc \\ 
  4 & Throttle & Brake & X-Acc & RPM & Heading \\ 
  5 & SW Acc & Brake & SW Angle & Heading & SW Vel \\ 
  6 & X-Acc & Gas & SW Vel & Brake & Speed \\ 
  7 & SW Vel & Brake & RPM & X-Acc & SW Angle \\ 
  8 & Lat Acc & Speed & SW Angle & Throttle & Gas \\ 
  9 & Brake & SW Angle & RPM & Speed & X-Acc \\ 
  10 & Brake & SW Vel & Speed & Heading & SW Acc \\ 
  11 & X-Acc & SW Acc & Brake & SW Vel & Throttle \\ 
  12 & X-Acc & RPM & SW Vel & Gas & Brake \\ 
\end{tabular}
\vspace{-5mm}
    \label{results3}
\end{table}

\section{Empirical Results}
We plot our results in Table \ref{results}. As shown, the accuracy varies significantly across the 12 different turns. For two-driver classification, predictions are between 55\% and 93.5\% accurate, with an average of 76.9\%. For five drivers, where a naive algorithm would only be correct 20\% of the time, our approach yields results between 33.2\% and 70\%, averaging 50.1\%. It is worth reiterating that these predictions come from only a single turn, just 8-10 seconds of sensor data. By examining the results more closely, we can see why there is such large variation across different turns. We look at the confusion matrix for turn \#8 in Figure \ref{fig:confusion1}, the turn with the lowest accuracy for $n = 5$. While predictions on drivers 1, 2, and 4 perform quite well (60\%, 57\%, and 39\% accuracy, respectively), those on drivers 3 (2\%) and 5 (8\%) achieve even worse-than-random results. This implies that drivers 3 and 5 did not have a clear ``style'' that was picked up by the random forest classifier. This could be due to the fact that these hard-to-classify drivers did not have consistent driving patterns, or that they were too similar to another driver. For an example of this latter case, consider the confusion between drivers 4 and 5 in Figure \ref{fig:turn6a}, where 
$\frac{2}{3}$ of driver 5's sessions were misclassified as coming from driver 4. To compare these results to an ideal case, where our classifier performs well, consider the confusion matrix for turn \#11, presented in Figure \ref{fig:confusion2}. Note that we take the five most frequent drivers \emph{at each turn}, so these 5 drivers are different than the ones in Figure \ref{fig:confusion1}. Turn \#11 yields strong classification results for all five drivers. This turn, shown in Figure \ref{fig:turn3}, is in a very rural area. Focusing on rural turns, we notice that the three best-performing turns, by a large margin, are the three which take place in rural locations: turns 10, 11, and 12. One explanation for this is that in less crowded areas, there are fewer potential obstacles such as pedestrians or other cars. Therefore, with more consistency across sessions, we are better able to discern driver's unique styles, which are less likely to be obscured by external conditions. 

\begin{figure}[]%!h]%!t?
\centering 
 \scalebox{1.0}{
  \subfigure[$n = 5$.]{\label{fig:turn6a}
$  	\begin{bmatrix}
       60 & 13 & 21 & 6 & 0     \\
       12 & 57 & 5 & 25 & 1     \\
       38 & 14 & 2 & 15 & 31   \\
       10 & 7 & 12 & 39 & 32   \\
       13 & 5 & 8 & 66 & 8  
     \end{bmatrix} $ 
     \vspace{1em}
  }
  
  \subfigure[$n=4$.]{\label{fig:turn6b}
$  	\begin{bmatrix}
       46 & 10 & 27 & 17     \\
       13 & 59 & 0 & 28     \\
       26 & 14 & 13 & 47   \\
       16 & 3 & 11 & 70    
     \end{bmatrix} $ 
     \vspace{1em}
  }}
  \subfigure[$n=3$.]{\label{fig:turn6c}
$  	\begin{bmatrix}
       69 & 8 & 23     \\
       19 & 69 & 12     \\
       40 & 44 & 16      
     \end{bmatrix} $ 
     \vspace{1em}
}
 \subfigure[$n=2$.]{\label{fig:turn6d}
$  	\begin{bmatrix}
       87 & 13     \\
       13 & 87   
     \end{bmatrix} $ 
     \vspace{1em}
  }
  \vspace{-1mm}
   \caption{Confusion matrix for turn with poor prediction accuracy.}
   \vspace{-3mm}
   \label{fig:confusion1}
\end{figure}

\begin{figure}[]%!h]%!t?
\centering 
  \subfigure[$n = 5$.]{\label{fig:turn7a}
$  	\begin{bmatrix}
       97 & 0 & 1 & 2 & 0     \\
       7 & 63 & 12 & 4 & 14     \\
       21 & 7 & 54 & 9 & 9   \\
       1 & 9 & 4 & 75 & 11   \\
       10 & 21 & 5 & 0 & 64  
     \end{bmatrix} $ 
     \vspace{1em}
  }
  \subfigure[$n=4$.]{\label{fig:turn7b}
$  	\begin{bmatrix}
       82 & 0 & 13 & 5     \\
       12 & 66 & 20 & 0     \\
       22 & 25 & 33 & 20   \\
       5 & 8 & 15 & 72   
     \end{bmatrix} $ 
     \vspace{1em}
  }
  \subfigure[$n=3$.]{\label{fig:turn7c}
$  	\begin{bmatrix}
       82 & 1 & 17     \\
       8 & 54 & 38     \\
       18 & 50 & 32      
     \end{bmatrix} $ 
     \vspace{1em}
}
 \subfigure[$n=2$.]{\label{fig:turn7d}
$  	\begin{bmatrix}
       95 & 5    \\
       8 & 92   
     \end{bmatrix} $ 
     \vspace{1em}
  }
  \vspace{-1mm}
   \caption{Confusion matrix for turn with good prediction accuracy.}
   \vspace{-5mm}
   \label{fig:confusion2}
\end{figure}

We next examine which sensors were the most ``important'' in helping our random forest distinguish between drivers. We rank the top 5 sensors for each turn's classifier in Table \ref{results3}. The results were very inconsistent across different turns. This implies that there is no one single ``indicator'' of driver identity, but rather that it is a combination of different factors, which manifests itself differently in different contexts. This also suggests that additional work is required to build more unified models, which can identify drivers more holistically, rather than depending on the specifics of each individual turn. However, while the sensor importance scores varied widely across the turns, several trends did emerge, such as forward acceleration (X-Acc), brake pedal, and steering wheel velocity being particularly relevant, especially at the turns with the highest prediction accuracy.
Additionally, we ran separate experiments comparing a random forest to other classification methods, given the set of features we used. Specifically, both multinomial logistic regression \cite{HS:04} and support vector machines (SVM) \cite{CV:95} yielded inferior classification results. Though results were comparable for $n=2$, as the pool of potential drivers increased, these two alternative approaches dropped off significantly compared to the random forest for the single-turn driver identification problem.

\xhdr{Predictions on Straightaways} 
To compare the turns in Table \ref{results} to relevant straightaways, we take 6 of these 12 turns where there are clear and well-defined straightaways immediately afterwards (i.e.\ excluding turns with another intersection right after it). This leaves 6 straightaways (3 urban, 3 rural), so we run our same classifier on these locations. Since the turns and straightaways happened only a few seconds apart, this ensures a fair comparison since it is unlikely that anything drastic happened to the driver or the road between these two intervals. We plot the results in Table \ref{results4}. As shown, we attain better accuracy at the turns than on straightaways. This is likely because, even though both cases have the same amount of raw data, the driver has much more to do during a turn, so the classifier is better able to discover the distinguishing characteristics. Additionally, just like the results for the turns, rural straightaways significantly outperform urban ones.

\begin{table}
    \caption{Prediction accuracy for straightaways immediately after the top turns from Table \ref{results}.}
    \scalebox{0.8}{
\begin{tabular}{ c | c | c | c | c | c }
  Corresponding & Straightaway & \multicolumn{4}{c}{Prediction Accuracy (\# of sessions per driver)} \\ 
  Turn ID & Location & $n$ = 2 & $n$ = 3 & $n$ = 4 & $n$ = 5  \\
  \hline			
  1 & Urban & 63.0\% (34) & 45.3\% (31) & 37.3\% (30) & 29.2\% (24) \\
  3 & Urban & 42.5\% (26) & 36.3\% (19) & 24.9\% (18) & 29.8\% (12) \\
  8 & Urban & 32.5\% (24) & 25.0\% (23) & 19.0\% (10) & 23.8\% (4) \\
  10 & Rural & 69.0\% (17) & 48.0\% (16) & 50.0\% (16) & 41.2\% (13) \\
  11 & Rural & 59.0\% (16) & 38.7\% (15) & 47.0\% (13) & 46.2\% (12) \\
  12 & Rural & 93.0\% (12) & 74.0\% (12) & 61.5\% (11) & 58.4\% (8) \\
  \hline
  & Urban average & 46.0\% & 35.5\% & 27.1\% & 27.6\% \\
  & Rural average & 73.7\% & 53.6\% & 52.8\% & 48.6\% \\
  & Total average & \textbf{59.8\%} & \textbf{44.6\%} & \textbf{40.0\%} & \textbf{38.1\%} \\
\end{tabular}
}
\vspace{-5mm}
    \label{results4}
\end{table}

\section{Conclusion and Future work}
In this paper, we have developed a method for predicting driver identity in automobiles using sensor data from a single turn. Our approach allows us to identify drivers by taking advantage of unique patterns of each individual's driving habits, which are evident even on a single turn. We test our algorithm on a dataset generated by Audi, consisting of real drivers on real roads in the area of Ingolstadt, Germany. We measure driver identification accuracy for each of the 12 most frequently made turns in the dataset, which comprise many different turn types. In these various conditions, we obtain average prediction accuracy of 76.9\% for two-driver classification, and 50.1\% for five drivers. With promising results and numerous potential applications, there are several directions for further exploration. We leave for future work the testing of our random forest-based method on longer time series, and a comparison on such datasets to the approach proposed by Miyajima et al. \cite{M:07}. Our classifier could also be extended with new spectral features or by incorporating more sensors. Additionally, a fusion approach combining several prediction models, including ours and alternative approaches such as Miyajima et al.'s, could lead to a classifier that is more stable than one that a single model can provide. Furthermore, our work could be used to analyze differences across turn types. For example, one could examine the classifier on major urban roads compared to highway entrances. This would show how the patterns used to identify each person may differ in distinct settings. This can be used to build ``driver profiles'' to model individual driving styles, an approach with much active research \cite{DGG:14, VLZ:14}. Overall, understanding drivers by analyzing sensor data has many practical benefits, and there are several direct applications and extensions of this work.


%\section*{Acknowledgments}
%The authors would like to thank Silei Xu for his help with this project. This research has been supported by Volkswagen Electronics Research Laboratory, AUDI AG, and Audi Electronics Venture.
%Silei, SIMPLEX?



%\bibliographystyle{IEEEtran}

\end{document}